\documentclass[%
preprint,superscriptaddress,
amsmath,amssymb,prd,aps,showpacs
floatfix,
]{revtex4-1}
\usepackage{graphics}
\usepackage{graphicx}
\usepackage{dcolumn}
\usepackage{bm}
\usepackage{mathrsfs}
\usepackage{amsmath}
\usepackage{amssymb}
\usepackage{float}
\usepackage{dcolumn}
\usepackage{multirow}
\usepackage{setspace}

\begin{document}


\title{Studying the bound state of the $K^-p$ system in the Bethe-Salpeter formalism}
\author{Chao Wang}
\email[Email: ]{chaowang@nwpu.edu.cn}
\affiliation{Center for Ecological and Environmental Sciences, Key Laboratory for Space Bioscience $\&$ Biotechnology, Northwestern Polytechnical University, Xi'an 710072, China}
\author{Liang-Liang Liu}
\email[Email: ]{liu06\_04@mail.bnu.edu.cn}
\affiliation{College of Physics and Information Engineering, Shanxi Normal University, Linfen 041004, China}
\affiliation{College of Nuclear Science and Technology,
Beijing Normal University, Beijing 100875, China}
\author{Xin-Heng Guo}
\email[Corresponding author, Email: ]{xhguo@bnu.edu.cn}
\affiliation{College of Nuclear Science and Technology,
Beijing Normal University, Beijing 100875, China}

\begin{abstract}
We study the $s$-wave kaon-nucleon bound state with the strangeness $S=-1$ in the Bethe-Salpeter formalism in the ladder and instantaneous approximations. We solve the Bethe-Salpeter equation of the bound state and obtain the Bethe-Salpeter amplitude. It is shown that the $K^-p$ bound state exists in this formalism. We also study the decay width of the bound state based on the Bethe-Salpeter techniques. The mass of this bound state is 1422\,MeV and its decay width is obviously smaller than that of $\Lambda(1405)$. These results indicate that there may be some other structures in the observed resonance.
\end{abstract}


\maketitle

\section{Introduction}

In the last decade, several nonconventional states were observed \cite{K.A.,Guo:2017jvc}. In the light baryon spectrum, one of such resonances is $\Lambda$(1405), which is just below the $\bar KN$ threshold and emerges in the meson-baryon scattering amplitude with  $I(J^P)=0(1/2^-)$ and strangeness $S=-1$ \cite{K.A.,Hyodo:2011ur}. It was discovered in the $\Sigma \pi$ invariant mass spectrum of the channel $K^-p\to\pi\pi\pi\Sigma$ \cite{Dalitz:1959dn,Dalitz:1960du,Alston:1961zzd}. Various experimental and theoretical investigations for $\Lambda(1405)$ have been preformed in recent years. The failure of the traditional quark model to interpret $\Lambda(1405)$ makes people consider exotic configurations such as the $\bar KN$ molecular structure \cite{Dalitz:1967fp,Bohm:1977mz}. This molecular configuration has been supported by recent lattice QCD results \cite{Hall:2014uca}.

Meanwhile, several studies propose that $\Lambda$(1405) has a two-pole structure and the spectrum in experiments exhibits one effective resonance shape \cite{Jido:2003cb}. Recently, the SIDDHARTA Collaboration has determined the energy shift and width of kaonic hydrogen, which provides a strong and direct constraint on the $\bar KN$ scattering amplitude \cite{Bazzi:2011zj,Bazzi:2012eq}. Based on this constraint, various studies confirm the two-pole structure of $\Lambda$(1405) \cite{Ikeda:2011pi,Ikeda:2012au,Mai:2012dt,Guo:2012vv,Mai:2014xna}. In this work, we will focus on the bound state below the $\bar KN$ threshold which might consist of the proton and antikaon.

The chiral perturbation theory (ChPT) is a useful effective field theory to deal with meson-meson (baryon) interactions at the low energy scale \cite{Gasser:1984gg,Ecker:1994gg}. A systematic and successful approach (the unitary chiral approach) combining ChPT and the unitarity condition of the scattering amplitude has been developed to describe the $K^-p$ scattering data \cite{Hyodo:2011ur,Oset:1997it}. In this paper, we will consider the Bethe-Salpeter equation method. The unitary chiral approach is based on the scattering theory. In this approach, the scattering amplitude can be obtained by regarding the two-body interaction obtained in chiral perturbation theory as the potential and solving the Lippmann-Schwinger equation \cite{Hyodo:2011ur,Oset:1997it}. The two-particle Bethe-Salpeter equation is derived from the relativistic quantum field theory. The basic concept is to relate the Bethe-Salpeter amplitude to the two-body propagator (four-point propagator) for which an integral equation can be derived from perturbation theory \cite{Feynman:1949hz,Salpeter:1951sz,DL}. In the unitary chiral approach, the bound state and resonance can be expressed as pole singularities in the scattering amplitude. The bound state and resonance poles appear in the first and second Riemann sheets, respectively \cite{Hyodo:2011ur,Oset:1997it}. In the Bethe-Salpeter technique, the bound state gives rise to a pole in the Fourier transform of the two-body propagator and we cannot get the continuum (or scattering) state by solving the homogeneous Bethe-Salpeter equation \cite{DL}. Both of the two approaches include a loop integration in momentum space, which might lead to an ultraviolet divergence. The unitary chiral approach deals with the divergence in the so-called on-shell factorization approach \cite{Hyodo:2011ur,Oset:1997it,Oller:1997ti}. In the Bethe-Salpeter technique, this divergence can be avoided which will be shown below \cite{Guo:1996jj,Guo:2007mm,Xie:2010zza,Feng:2011zzb}. Besides, we solve Bethe-Salpeter amplitude from the Bethe-Salpeter equation, while the scattering amplitude is solved in the unitary chiral approach. The dependence of the Bethe-Salpeter amplitude on the momentum transform reveals the structure of the bound states \cite{Guo:1996jj,Guo:2007mm,Xie:2010zza,Feng:2011zzb}.

The Bethe-Salpeter techniques were developed by Feynman, Salpeter, and Bethe \cite{Feynman:1949hz,Salpeter:1951sz,DL}. It has been applied to theoretical studies concerning heavy baryons and molecular bound states \cite{Guo:1996jj,Guo:2007mm,Xie:2010zza,Feng:2011zzb}. In previous studies, the possible bound states of $K\bar K$, $DK$, and $B\bar K$ have been investigated in the Bethe-Salpeter formalism with the kernel introduced by ChPT in the ladder and instantaneous approximations \cite{Guo:2007mm,Xie:2010zza,Feng:2011zzb}. We will try to study the $K^-p$ bound state in this framework in the present paper. We will take these two approximations into account and consider the interaction kernel provided by the leading order of ChPT. We will investigate whether the bound state exists or not and study its decay in this picture. We will also discuss the extent to which the $K^-p$ component contributes to the observed $\Lambda(1405)$ resonance.

The remainder of this paper is organized as follows. In Sec.~II, we derive the Bethe-Salpeter equation for the $K^-p$ system in detail and present the normalization condition of the Bethe-Salpeter amplitude. In Sec.~III, the decay of the $K^-p$ bound state to $\Sigma^+\pi^-$ is discussed. The numerical results are presented in Sec.~IV. In the last section, we give a summary and some discussions.

\section{Bethe-Salpeter formalism for the bound state containing the proton and antikaon}

\subsection{Bethe-Salpeter equation for the $K^-p$ system }
In this section, we will derive the Bethe-Salpeter equation for the $K^-p$ system. We assume that the bound state exists and its mass is $M$. We denote it by $\Lambda^*$. In this picture, the Bethe-Salpeter amplitude can be defined as \cite{DL,Guo:1996jj,Guo:2007mm,Xie:2010zza,Feng:2011zzb,Zhang:2013gqa,Liu:2015qfa,Liu:2016wzh}:
\begin{eqnarray}
\chi(x_1,x_2,P)&=&\langle 0|T\psi(x_1) \phi(x_2)|\Lambda^*\rangle,
\end{eqnarray}
with $\psi(x_{1})$ and $\phi(x_{2})$ being field operators of the proton and $K^-$, respectively, and $P$ being the momentum of the system. In the momentum space, the Bethe-Salpeter amplitude, $\chi_{P}(p)$, is related to $\chi(x_1,x_2,P)$ through the following equation \cite{DL}:
\begin{eqnarray}
\chi(x_{1},x_{2},P)=\mathrm e^{\mathrm iPX}\int\frac{\mathrm{d}^{4}p}{(2\pi)^4}\chi_{P}(p)\mathrm e^{\mathrm ipx},
\end{eqnarray}
where $p$ and $x(=x_{1}-x_{2})$ are the relative momentum and the relative coordinate of two constituents, respectively, and $X$ is the center of mass coordinate which is defined as $X=\lambda_{1}x_{1}+\lambda_{2}x_{2}$, where $\lambda_{1}=\frac{m_{1}}{m_{1}+m_{2}}$, $\lambda_{2}=\frac{m_{2}}{m_{1}+m_{2}}$, with $m_{1}$ and $m_{2}$ being the masses of the proton and the $K^-$ meson, respectively. The momentum of the proton is $p_{1}=\lambda_{1}P+p$ and that of $K^-$ is $p_{2}=\lambda_{2}P-p$. The derivation of the Bethe-Salpeter formalism for the two fermion systems can be found in the textbook \cite{DL}. In the same way, one can prove that the form of the Bethe-Salpeter equation is still valid for the fermion and scalar object system \cite{Guo:1996jj}. Therefore, the Bethe-Salpeter amplitude in our case satisfies the follow homogeneous integral equation \cite{DL,Guo:1996jj,Guo:2007mm,Xie:2010zza,Feng:2011zzb,Zhang:2013gqa,Liu:2015qfa,Liu:2016wzh}:
\begin{eqnarray}\label{BSe1.1}
\chi_{P}(p)&=&s_{F}(\lambda_{1}P+p)\int\frac{\mathrm d^{4}q}{(2\pi)^4}K(P,p,q) \chi_{P}(q)s_{B}(\lambda_{2}P-p),
\end{eqnarray}
where $s_F$ and $s_B$ are propagators of the proton and $K^-$, respectively. We also define the relative longitudinal momentum $p_l(=v\cdot p)$ and transverse momentum $p_t[=p-(v\cdot p)v]$ with $v(=P/M)$ being the 4-velocity of the bound state, and $K(P,p,q)$ is the interaction kernel that can be described by the sum of all the irreducible graphs which cannot be split as into two pieces by cutting two-particle lines as defined in Ref.~\cite{DL}. For the propagators, we have
\begin{eqnarray}\label{pg1.1}
&& s_F(\lambda_1P+p)=\frac{\mathrm i \left [(\lambda_1M+p_l) v\!\!\!/+p\!\!\!/_t+m_1\right]}{(\lambda_1M+p_l-\omega_1+\mathrm i \epsilon)(\lambda_1M+p_l+\omega_1-\mathrm i \epsilon)},\nonumber\\[9pt]
&&s_B(\lambda_2P-p)= \frac{\mathrm i}{(\lambda_2M-p_l-\omega_2+\mathrm i \epsilon)(\lambda_2M-p_l+\omega_2-\mathrm i \epsilon)},
\end{eqnarray}
with $\omega_{1(2)}=\sqrt{m_{1(2)}^2-p_t^2}$.

In general, considering $v\!\!\!/u(v, s) = u(v, s)$ , $\chi_P(p)$ can be written as \cite{Zhang:2013gqa,Liu:2015qfa,Liu:2016wzh}
\begin{eqnarray}
\chi_{P}(p)=(g_1+g_2\gamma_5+g_3\gamma_5p\!\!\!/_t+g_4p\!\!\!/_t+g_5\sigma_{\mu \nu }\varepsilon^{\mu \nu\alpha\beta}p_{t\alpha}v_{\beta})u(v,s),
\end{eqnarray}
where $u(v,s)$ is the spinor of the bound state with helicity $s$ and $g_i$($i=1,2\cdots5$) are Lorentz-scalar functions. With the constraints imposed by parity and Lorentz transformations, it is easy to prove that $\chi_{P}(p)$ can be simplified as \cite{Zhang:2013gqa,Liu:2015qfa,Liu:2016wzh}
\begin{eqnarray}\label{wf1.1}
\chi_{P}(p)=[f_1(p)+f_2(p)p\!\!\!/_t]u(v,s),
\end{eqnarray}
in which $f_1(p)$ and $f_2(p)$ are two independent Lorentz-scalar functions of $p$.

\begin{figure}[tb]
\begin{center}
\scalebox{0.78}[0.78]{\includegraphics[193,603][402,721]{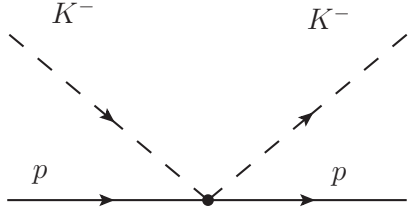}}
\caption{Feynman diagram for the Weinberg-Tomozawa interaction in ChPT.}\label{WT}
\end{center}
\end{figure}

The Bethe-Salpeter equation can be treated in the so-called ladder approximation \cite{DL}. In this approximation, $K(P,p,q)$ is replaced by its lowest-order value \cite{DL}. Here we adapt ChPT to describe the s-wave meson-baryon interaction. For the $s$-wave amplitude, the most important piece is the Weinberg-Tomozawa contact interaction at the lowest order $\mathcal O(p)$ of ChPT \cite{Hyodo:2011ur,Weingerg:1996}. According to ChPT, the Lagrangian for the Weinberg-Tomozawa contact interaction as shown in Fig.~\ref{WT} is \cite{Hyodo:2011ur,Ecker:1994gg,Oset:1997it}
\begin{eqnarray}\label{chpt}
\mathcal L=\bar\psi \mathrm i \gamma^\mu\frac{C_{\mathrm {if}}}{4f^2}(\phi\partial_{\mu} \phi- \partial_{\mu} \phi \phi)\psi,
\end{eqnarray}
where $ C_{\mathrm {if}}$ is the isospin coefficient (i and f represent the initial and final states, respectively) and $f$ corresponds to the meson decay constant in the chiral limit at the tree level. Using this interaction term, we obtain the formalism of $K(P,p,q)$ in terms of $p_l$ and $p_t$:
\begin{eqnarray}\label{kernel1.1}
K(P,p,q)&=&-\mathrm i C_{\mathrm {if}}\frac{1}{4f^2}(p\!\!\!/_2+q\!\!\!/_2)\nonumber\\
&=&-\mathrm i C_{\mathrm {if}}\frac{1}{4f^2}\left[2(\lambda_2 M-p_l)v\!\!\!/-(p\!\!\!/_t+q\!\!\!/_t)\right],
\end{eqnarray}
where we use the covariant instantaneous approximation, $p_l=q_l$, in the last line .

Now we substitute Eqs.~(\ref{pg1.1}), (\ref{wf1.1}) and (\ref{kernel1.1}) into the Bethe-Salpeter equation (\ref{BSe1.1}) and obtain the coupled integral equations about $f_1¡¡(p_l,\,{p}_t)$ and $f_2¡¡(p_l,\,{p}_t)$:
\begin{eqnarray}\label{BSf11.1}
&&4f_1(p_l,\, {p}_t)=\frac{\mathrm i C_{\mathrm {if}}}{f^2}\int\frac{\mathrm d^4 q}{(2\pi)^4}\nonumber\\[8pt]
&&\cdot \Bigg[ \frac{\left[2(\lambda_1M+p_l)(\lambda_2M-p_l)-p_t^2-p_t\cdot q_t\right]\cdot f_1(q_l,\,{q}_t)}{(\lambda_1M+p_l-\omega_1+\mathrm i \epsilon)(\lambda_1M+p_l+\omega_1-\mathrm i  \epsilon)(\lambda_2M-p_l-\omega_2+\mathrm i \epsilon)(\lambda_2M-p_l+\omega_2-\mathrm i \epsilon)}\nonumber\\[8pt]
&&\qquad-\frac{m_1(p_t\cdot q_t +q_t^2)\cdot f_2(q_l,\,{q}_t)}{(\lambda_1M+p_l-\omega_1+\mathrm i \epsilon)(\lambda_1M+p_l+\omega_1-\mathrm i \epsilon)(\lambda_2M-p_l-\omega_2+\mathrm i \epsilon)(\lambda_2M-p_l+\omega_2-\mathrm i \epsilon)}
\Bigg],\nonumber\\[3pt]
\end{eqnarray}
\begin{eqnarray}\label{BSf21.1}
&&4p_t^2f_2(p_l,\, {p}_t)=\frac{\mathrm i C_{\mathrm {if}}}{f^2}\int\frac{\mathrm d^4 q}{(2\pi)^4}\nonumber\\[8pt]
&&\cdot \Bigg[ \frac{-m_1\left(p_t^2+p_t\cdot q_t\right)\cdot f_1(q_l,\,{q}_t)}{(\lambda_1M+p_l-\omega_1+\mathrm i \epsilon)(\lambda_1M+p_l+\omega_1-\mathrm i \epsilon)(\lambda_2M-p_l-\omega_2+\mathrm i \epsilon)(\lambda_2M-p_l+\omega_2-\mathrm i \epsilon)}\nonumber\\[8pt]
&&\quad+\frac{\left[2(\lambda_1M+p_l)(\lambda_2M-p_l)p_t\cdot q_t-p_t^2p_t\cdot q_t-p_t^2q_t^2\right]f_2(q_l,\,{q}_t)}{(\lambda_1M+p_l-\omega_1+\mathrm i \epsilon)(\lambda_1M+p_l+\omega_1-\mathrm i \epsilon)(\lambda_2M-p_l-\omega_2+\mathrm i \epsilon)(\lambda_2M-p_l+\omega_2-\mathrm i \epsilon)}\Bigg].\nonumber\\[3pt]
\end{eqnarray}
We define the functions $\tilde{f}_{1(2)}(p_t)=\int \frac{\mathrm d p_l}{2\pi}f_{1(2)}(p_l,\, p_t)$. In the $\Lambda^*$ rest frame, one has $p_{t}=(0,\,-\vec{ p_t})$ and $|p_t|=|\vec{p}_t|$. Performing the integration over $p_l$ on both sides through the residue theorem, we find that $\tilde{f}_{1(2)}(\vec p_t)$ satisfy the coupled integral equations as follows:
\begin{eqnarray}\label{BSbf11.1}
&&\tilde{f}_{1}(\vec{p}_t)=\frac{ C_{\mathrm {if}}}{4f^2}\int\frac{\mathrm d^3 \vec{q}_t}{(2\pi)^3}\nonumber\\[9pt]
&&\cdot\Bigg[\bigg(\frac{2(2\lambda_1M+\omega_1)(M+\omega_1)-|\vec{p}_t|^2-\vec{p}_t\cdot\vec{q}_t}{2\omega_1[(M+\omega_1)^2-\omega_2^2]}+\frac{2\omega_2(M-\omega_2)+|\vec{p}_t|^2+\vec{p}_t\cdot\vec{q}_t}{2\omega_2[(M-\omega_2)^2-\omega_1^2]}\bigg)\cdot\tilde{f}_{1}(\vec{q}_t)\nonumber\\[9pt]
&&\qquad\qquad-m_1\cdot\bigg(\frac{|\vec{q}_t|^2+\vec{p}_t\cdot\vec{q}_t}{2\omega_1[(M+\omega_1)^2-\omega_2^2]}-\frac{|\vec{q}_t|^2+\vec{p}_t\cdot\vec{q}_t}{2\omega_2[(M-\omega_2)^2-\omega_1^2]}\bigg)\cdot\tilde{f}_{2}(\vec{q}_t)
\Bigg],
\end{eqnarray}
\begin{eqnarray}\label{BSbf21.1}
\tilde{f}_{2}(\vec{p}_t)&=&\frac{C_{\mathrm {if}}}{4(-|\vec{p}_t|^2)f^2}\int\frac{\mathrm d^3 \vec{q}_t}{(2\pi)^3}\cdot\Bigg[\bigg(\frac{m_1\left(|\vec{p}_t|^2+\vec{p}_t\cdot\vec{q}_t\right)}{2\omega_1[(M+\omega_1)^2-\omega_2^2]}-\frac{m_1\left(|\vec{p}_t|^2+\vec{p}_t\cdot\vec{q}_t\right)}{2\omega_2[(M-\omega_2)^2-\omega_1^2]}\bigg)\cdot\tilde{f}_{1}(\vec{q}_t)\nonumber\\[9pt]
&&\qquad-\bigg(\frac{2(2\lambda_1M+\omega_1)(M+\omega_1)\vec{p}_t\cdot\vec{q}_t-|\vec p_t|^2\vec{p}_t\cdot\vec{q}_t-|\vec p_t|^2|\vec q_t|^2}{2\omega_1[(M+\omega_1)^2-\omega_2^2]}\nonumber\\[9pt]
&&\qquad\qquad\qquad+\frac{2\omega_2(M-\omega_2)\vec{p}_t\cdot\vec{q}_t+|\vec p_t|^2\vec{p}_t\cdot\vec{q}_t+|\vec p_t|^2|\vec q_t|^2}{2\omega_2[(M-\omega_2)^2-\omega_1^2]}\bigg)\cdot\tilde{f}_{2}(\vec{q}_t)
\Bigg],
\end{eqnarray}
with $\omega_{1(2)}=\sqrt{m_{1(2)}^2+|\vec p_t|^2}$. These two equations involve the integrations of $q_t$. They look like divergent integrations since $q_t$ varies from 0 to $+\infty$. However, the Bethe-Salpeter amplitudes ($f_1$ and $f_2$) decrease to zero rapidly at the large momentum transfer and thus there is no divergence in practice \cite{DL,Guo:1996jj,Guo:2007mm,Xie:2010zza,Feng:2011zzb,Feng:2011zzb}.

\subsection{Normalization condition of the Bethe-Salpeter amplitude}
In Eqs.~(\ref{BSf11.1}) and (\ref{BSf21.1}), we leave the normalization of $\tilde{f}_1$ and $\tilde{f}_2$ undetermined. Following Ref.\cite{DL}, the normalization condition for the Bethe-Salpeter equation can be written as
\begin{eqnarray}\label{nc1.2}
&&\frac{\mathrm i}{(2 \pi)^4}\int\mathrm d^4 p \mathrm d^4 q \bar \chi_P(p)\frac{\partial }{\partial P_0}[I(p,\,q,\,P)+K(p,\,q,\,P)]\chi_P(q)=2P_0,
\end{eqnarray}
where $I(p,\,q,\,P)$ is the inverse of the four-point propagator
\begin{eqnarray}\label{nc1.2}
I(p,\,q,\,P)=\delta^{(4)}(p-q)[s_F(\lambda_1P + p)]^{-1}[s_B(\lambda_2P - p)]^{-1}.
\end{eqnarray}
One can recast the normalization condition for the Bethe-Salpeter amplitude into the form \cite{Liu:2015qfa}
\begin{eqnarray}\label{ncc1.2}
&&-\int\frac{\mathrm{d}^4p}{(2\pi)^4}\Big\{\mathrm{Tr}[\alpha_P(p)\beta_P(p)s_F(p_1)(\lambda_1\varepsilon\!\!\!/)s_F(p_1)s_B(p_2)]\nonumber\\
&&\qquad\qquad\quad+\mathrm{Tr}[\alpha_P(p)\beta_P(p)(2\lambda_2p_2\cdot\varepsilon)s_F(p_1)s_B(p_2)s_B(p_2)]\Big\}=2P_0,
\end{eqnarray}
where $\varepsilon=(1,\,0,\,0,\,0)$ and $\alpha_P(p_l,\,p_t)$[$\beta_P(p_l,\,p_t)$] is the transverse projection
of the Bethe-Salpeter amplitude given by
\begin{eqnarray}\label{aalphabeta1.2}
\alpha_P(p_l,\,{p_t})&=&-\mathrm{i}s_F(p_1)^{-1}\chi_P(p_l,{p_t})s_B(p_2)^{-1},\nonumber\\
\beta_P(p_l,\,{p_t})&=&-\mathrm{i}s_B(p_2)^{-1}\bar{\chi}_P(p_l,{p_t})s_F(p_1)^{-1}.
\end{eqnarray}
Substituting Eq.~(\ref{BSe1.1}) into the above equations, we obtain
\begin{eqnarray}\label{alphabeta1.2}
\alpha_P(p_l,\,p_t)&=&[\tilde{h}_1( p_t)+p\!\!\!/_t\tilde{h}_2( p_t)]u(v,s),\nonumber\\
\beta_P(p_l,\,p_t)&=&\bar u(v,s)[\tilde{h}_1( p_t)+p\!\!\!/_t\tilde{h}_2(p_t)],
\end{eqnarray}
with
\begin{eqnarray}\label{h1h21.2}
\tilde{h}_1(p_t)&=&\int\frac{\mathrm{d}^3 q_t}{(2\pi)^3}\frac{C_{\mathrm {if}}( p_t\cdot q_t+|{q}_t|^2)\tilde f_2( q_t)}{4f^2|{p}_t|^2},\nonumber\\
\tilde{h}_2( p_t)&=&\int\frac{\mathrm{d}^3 q_t}{(2\pi)^3}\frac{C_{\mathrm {if}}(p_t\cdot q_t+|{q}_t|^2)\tilde f_1( q_t)}{4f^2}.
\end{eqnarray}
Then, we substitute Eqs.~(\ref{alphabeta1.2}) and (\ref{h1h21.2}) into Eq.~(\ref{ncc1.2}) and integrate out the relative longitudinal momentum $q_l$. In the $\Lambda^*$ rest frame, the normalization condition can be written in the following form
\begin{eqnarray}
&&-\int\frac{\mathrm{d}^3\vec p_t}{2(2\pi)^3}\frac{4M}{(M+\omega_1-\omega_2)^2}\Bigg\{\frac{\lambda_1}{2\omega_2(M-\omega_1-\omega_2)^2}\Big[-4\tilde{h}_1(\vec p_t)\tilde{h}_2(\vec p_t)\vec p_t^2(\lambda_2M-\omega_2)\nonumber\\[5pt]
&&\qquad\qquad\quad -\tilde{h}_1^2(\vec p_t)\big(\lambda_2^2M^2+m_1^2-2\lambda_2M\omega_2+2\lambda_2Mm_1-2m_1\omega_2+\omega_2^2-\vec p_t^2 \big)\nonumber\\[5pt]
&&\qquad\qquad\quad +\tilde{h}_2^2(\vec p_t)\vec p_t^2 \big(\lambda_2^2M^2-2\lambda_2M m_1+m_1^2+2m_1\omega_2+\omega_2^2-\vec p_t^2\big)\Big]\nonumber\\[5pt]
&&\qquad+\frac{\lambda_2(M+\lambda_1\lambda_2M-\omega_1)}{\omega_1(M+\omega_1+\omega_2)^2}\Big[ \tilde{h}_1^2(\vec p_t)\big( m_1-\lambda_1M-\omega_1\big)+\tilde{h}_1(\vec p_t)\tilde{h}_2(\vec p_t)\vec p_t^2  \nonumber\\[5pt]
&&\qquad\qquad\quad+\tilde{h}_2^2(\vec p_t)\vec p_t^2\big(\lambda_1M+\omega_1+m_1 \big)\Big]\Bigg\}
=2M.
\end{eqnarray}

\section{The decay width of the $K^-p$ bound state}

\begin{figure}[bt]
\begin{center}
\includegraphics[width=0.6\textwidth]{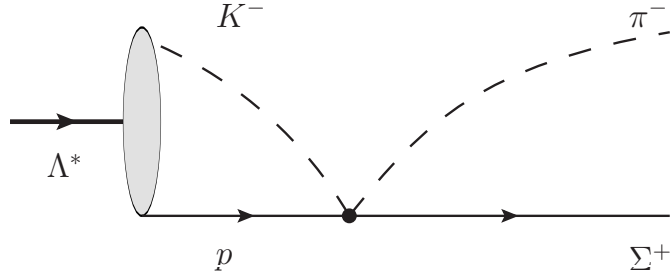}
\caption{Feynman diagram for the decay $\Lambda^*\to \Sigma^+\pi^-$.\label{fig3}}
\end{center}
\end{figure}

In this section, we will proceed to apply the Bethe-Salpeter technique to derive the decay width of the $K^-p$ bound state. According to the experiments, $\Lambda(1405)$ exclusively decays into $\Sigma\pi(I=0)$. In fact, Dalitz and Deloff analyzed the $\Sigma^+\pi^-$ spectrum to extract the mass and width of $\Lambda(1405)$ \cite{Dalitz:1991sq}. This clear spectrum is frequently shown as a representative of the $\Lambda(1405)$ spectrum and used for the input of theoretical models \cite{Hyodo:2011ur}. Therefore, we will study the decay width of the $K^-p$ bound state into the above final state. As shown in Fig.~\ref{fig3}, the relevant interaction vertex is given in Eq.~(\ref{chpt}) at the lowest order $\mathcal O (p)$ through ChPT \cite{Hyodo:2011ur,Oset:1997it}. We define $p_a[=(E_a,\,-\vec p_a)]$ and $p_b[=(E_b,\,-\vec p_b)]$ to be the momenta of $\Sigma^+$ and $\pi^-$, respectively. $p^\prime(=\lambda^\prime_2 p_a-\lambda^\prime_1 p_b)$ is defined as the relative momentum of $\Sigma^+$ and $\pi^-$ where $\lambda^\prime_{1}=\frac{m_{a}}{m_{a}+m_{b}}$, $\lambda^\prime_{2}=\frac{m_{b}}{m_{a}+m_{b}}$, with $m_{a}$ and $m_{b}$ being the masses of $\Sigma^+$ and $\pi^-$, respectively, and $p^\prime_l(=v\cdot p^\prime)$ is the relative longitudinal momentum of $\Sigma^+$ and $\pi^-$. According to the kinematics of the two-body decay, in the rest frame of the bound state one has
\begin{eqnarray}
E_b&=&\frac{M^2-m_a^2+m_b^2}{2M}, \qquad E_a=\frac{M^2-m_b^2+m_a^2}{2M},\nonumber\\[5pt]
|\vec p_a|&=&|\vec p_b|=\frac{\sqrt{(M^2-(m_a+m_b)^2)(M^2-(m_a-m_b)^2)}}{2M}.
\end{eqnarray}
The differential decay width reads
\begin{eqnarray}
\mathrm d \Gamma=\frac{1}{32 \pi^2}|\mathcal M|^2\frac{|\vec p_a|}{M^2}\mathrm d \Omega,
\end{eqnarray}
where $\Omega$ is the solid angle of $\Sigma^+$.

\begin{figure}[bt]
\begin{center}
\includegraphics[width=0.5\textwidth]{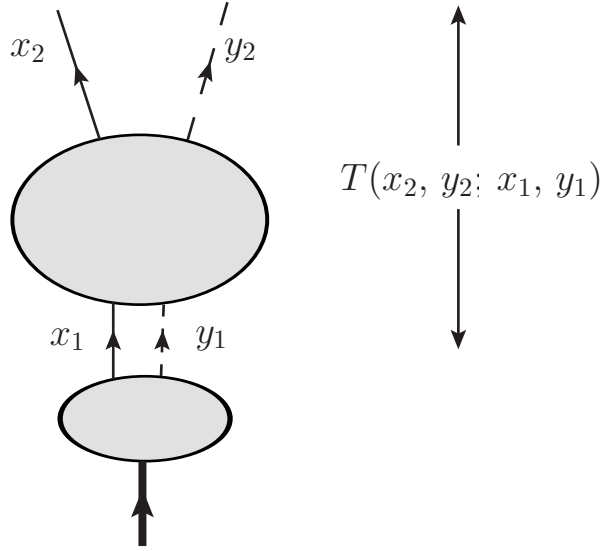}
\caption{The truncated Bethe-Salpeter irreducible part $T(x_2,\,y_2;\,x_1,\,y_1)$.\label{Tx1y1}}
\end{center}
\end{figure}

Next, we will present the amplitude based on the techniques in the textbook \cite{DL}. According to the LSZ reduction formula, the S-matrix element for this process is
\begin{eqnarray}\label{lsz0}
\langle\Sigma(p_a)\pi(p_b)|\Lambda^*(P)\rangle_H&=&-\text{i}\int\text{d}^4x_2\text{d}^4y_2\mathrm{e}^{\mathrm i p_ax_2} \mathrm{e}^{\mathrm i p_by_2} \bar{u}_\Sigma(p_a) \nonumber\\
&& \qquad\cdot (m_b^2-\partial_{y_2}^2)(\mathrm i \partial\!\!\!/_{x_2}-m_a)  \langle 0|\psi(x_2)\phi(y_2)|\Lambda^*(P)\rangle_I,
\end{eqnarray}
where $I$ and $H$ represent the interaction and Heisenberg pictures, respectively.
Following Eq.~9(77) in Ref.~\cite{DL}, one has
\begin{eqnarray}\label{T0}
\langle 0|\psi(x_2)\phi(y_2)|\Lambda^*(P)\rangle_I=-\int\mathrm d^4 x_1\mathrm d^4 y_1 T(x_2,\,y_2;x_1,\,y_1)\chi_P(x_1,\,y_1),
\end{eqnarray}
in which $T(x_2,\,y_2;\,x_1,\,y_1)$ is defined to be the truncated Bethe-Salpeter irreducible part. By 'truncation', we mean the removal of the two propagators corresponding to the incoming lines as shown in Fig.~\ref{Tx1y1}. $T(x_2,\,y_2;\,x_1,\,y_1)$ can be evaluated perturbatively:
\begin{eqnarray}\label{T}
&&T(x_2,\,y_2;\,x_1,\,y_1)\nonumber\\
&=&\frac{C_{\mathrm {if}}}{4f^2}\int \mathrm d^4 x_i s_1(x_i-x_2)s_2(x_i-y_2)\gamma^\mu(\partial_{x_i}-\overleftarrow{\partial}_{x_i})\delta^{(4)}(x_1-x_i)\delta^{(4)}(y_1-x_i)\nonumber\\[5pt]
&=&\frac{C_{\mathrm {if}}}{4f^2}\int \mathrm d^4 x_i\int \frac{\mathrm d^4 p_A}{(2\pi)^4}\mathrm e ^{\mathrm i p_A(x_i-x_2)}(\mathrm i p\!\!\!/_A-m_a)^{-1} \int \frac{\mathrm d^4 p_B}{(2\pi)^4}\mathrm e ^{\mathrm i p_B(x_i-y_2)}( p_B^2-m_b^2)^{-1}\gamma^\mu(\partial_{x_i}-\overleftarrow{\partial}_{x_i}) \nonumber\\[5pt]
&&\hspace{3.5cm} \times \delta^{(4)}(x_1-x_i)\delta^{(4)}(y_1-x_i).
\end{eqnarray}
Substituting Eqs.~(\ref{T0}) and (\ref{T}) into Eq.~(\ref{lsz0}), we have
\begin{eqnarray}\label{lsz}
&&\langle\Sigma(p_a)\pi(p_b)|\Lambda^*(P)\rangle_H\nonumber\\
&=&\text{i}\frac{C_{\mathrm{if}}}{4f^2}\int\text{d}^4x_2\text{d}^4y_2\mathrm{e}^{\mathrm i p_a x_2}\mathrm{e}^{\mathrm i p_by_2}\bar{u}(p_a)(\mathrm i \partial\!\!\!/_{x_2}-m_a)(m_b^2-\partial_{y_2}^2)\int\mathrm d^4 x_1\mathrm d^4 y_1 \int \mathrm d^4 x_i \delta^{(4)}(x_1-x_i)\delta^{(4)}(y_1-x_i)\nonumber\\
&&\qquad\cdot\int \frac{\mathrm d^4 p_A}{(2\pi)^4}\mathrm e ^{\mathrm i p_A(x_i-x_2)}(\mathrm i p\!\!\!/_A-m_a)^{-1}\int \frac{\mathrm d^4 p_B}{(2\pi^4)}\mathrm e ^{\mathrm i p_B(x_i-y_2)}( p_B^2-m_b^2)^{-1}\gamma^\mu(\partial_{x_i}-\overleftarrow{\partial}_{x_i}) \chi_P(x_1,\,y_1)\nonumber\\[4pt]
&=&\mathrm{i}\frac{C_{\mathrm {if}}}{4f^2}\bar{u}(p_a)\int\mathrm{d}^4x_1\mathrm{d}^4y_1\int\mathrm{d}^4x_i\mathrm{e}^{\mathrm{i}p_a x_i}\mathrm{e}^{\mathrm{i}p_b x_i}(p\!\!\!/_2+p\!\!\!/_b)\delta^{(4)}(x_1-x_i)\delta^{(4)}(y_1-x_i)\chi_P(x_1,\,y_1)\nonumber\\[4pt]
&=&\mathrm{i}(2\pi)^4\delta(P-p_a-p_b)\frac{C_{\mathrm {if}}}{4f^2}\int\frac{\mathrm{d}^4p}{(2\pi)^4}\bar{u}(p_a)(p\!\!\!/_2+p\!\!\!/_b)\chi_P(p),
\end{eqnarray}
where we use the relation $\chi_P(x_i,x_i)=\mathrm{e}^{\mathrm{i}Px_i}\int \frac{\mathrm{d}^4 p}{(2\pi)^4} \chi_P(p)$. According to  Eqs.~(\ref{wf1.1}) and (\ref{lsz}), the amplitude of the $\Lambda^*\to \Sigma^+\pi^-$ process is
\begin{eqnarray}
\mathcal M=\mathrm i\frac{C_{\mathrm {if}}}{4f^2}\int\frac{\mathrm{d}^3 p_t}{(2\pi)^3}\bar{u}_\Sigma(p_a)[(\lambda_2M-p_l^\prime) v\!\!\!/-p\!\!\!/_t+p\!\!\!/_b] \cdot(\tilde{f}_1( p_t)+\tilde{f}_2(p_t)p\!\!\!/_t)u_\Lambda(P),
\end{eqnarray}
in which we have considered the instantaneous approximation, $p_l= p_l^\prime$, again.

In the $\Lambda^*$ rest frame, averaging over the spins of the initial state and summing over the spins of the final state, the unpolarized decay width of $\Lambda^*$ is
\begin{eqnarray}
\Gamma&=  &\frac{|\vec p_a|}{32\pi^5M^2} \frac{C^2_{\mathrm {if}}}{16f^4}\int \mathrm{d}|\vec q_t||\vec q_t|^2\int \mathrm{d}|\vec p_t||\vec p_t|^2 \nonumber\\
&&\cdot\Big \{4M\tilde{f}_1(|\vec p_t|)\tilde{f}_1(|\vec q_t|)\big[(2E_b+2p_l^\prime)(p_a\cdot p_b)-E_a m_b^2+E_ap_l^{\prime2}+2E_b m_a p_l^\prime+m_am_b^2+m_a p_l^{\prime2}\big]\nonumber\\
&&\qquad+4\tilde{f}_2(|\vec p_t|)\tilde{f}_2(|\vec q_t|)|\vec p_t|^2|\vec q_t|^2(E_a+m_a)+4\tilde{f}_1( |\vec p_t|)\tilde{f}_2(|\vec q_t|)|\vec q_t|^2\big[(p_a\cdot p_b)+E_a p_l^\prime\nonumber\\[2pt]
&&\qquad\quad+E_bm_a+m_a p_l^\prime \big]+4\tilde{f}_2(\vec p_t)\tilde{f}_1(|\vec q_t|)|\vec p_t|^2\big[(p_a\cdot p_b)+E_a p_l^\prime+E_bm_a+m_a p_l^\prime \big]\Big \},
\end{eqnarray}
with $p_l^\prime=\lambda^\prime_2E_a-\lambda^\prime_1E_b$.

\section{Numerical result}

\begin{figure}[bt]
\begin{center}
\scalebox{1}[1]{\includegraphics[width=0.8\textwidth, viewport=445pt 8pt 797pt 375pt,clip]{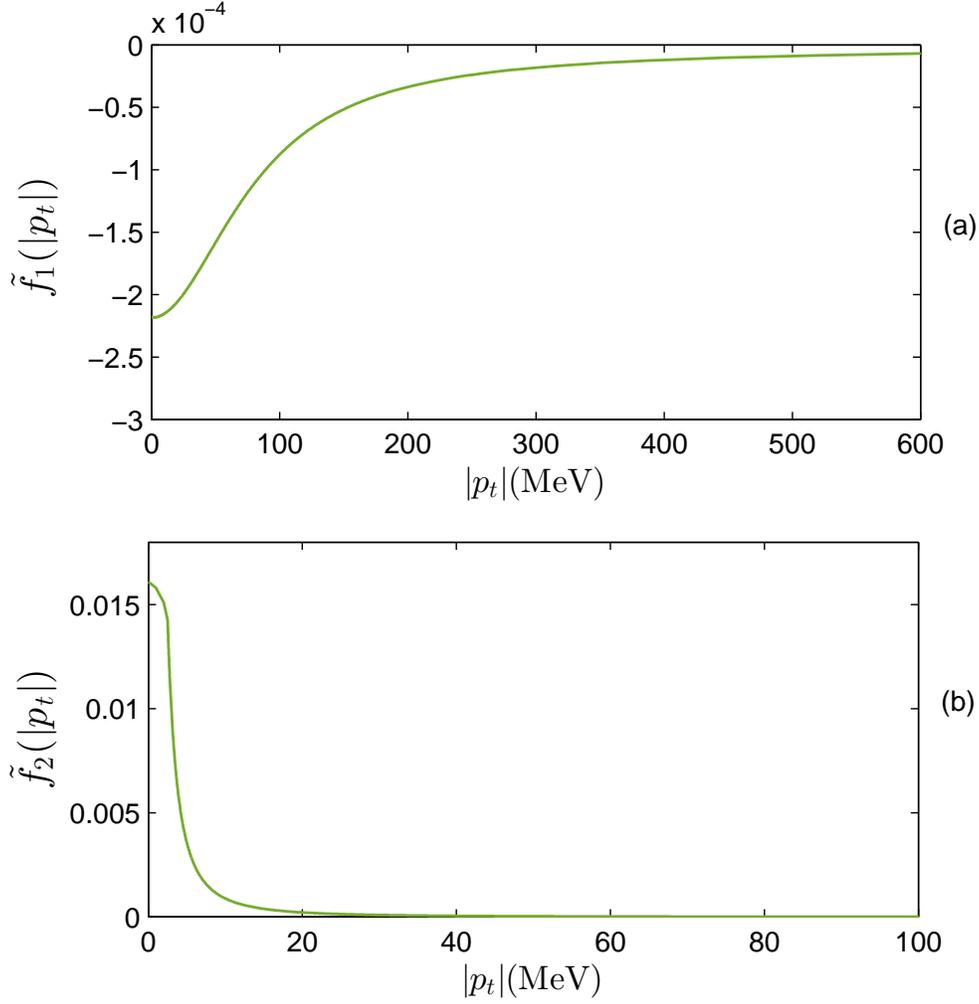}}
\caption{Numerical result for the normalized Bethe-Salpeter amplitude of the bound state. The units of $\tilde{f_1}$ and $\tilde{f_2}$ are 1 and $\mathrm{MeV}^{-1}$, respectively.}\label{figBSWave}
\end{center}
\end{figure}

In this part, we will solve the Bethe-Salpeter equation numerically and try to search for the possible solution of the $K^-p$ bound state. To find out the bound state in this system, one only needs to solve the homogeneous Bethe-Salpeter equation. One solution corresponds to a possible bound state. Since the Bethe-Selpeter amplitude for the ground state is in fact rotationally invariant, $\tilde{f}_{1(2)}$ depends only on $|p_t|$. Generally, $|p_t|$ varies from 0 to $+\infty$ and $\tilde{f}_{1(2)}$ would decrease to zero when $| p_t|\to +\infty$. We replace $| p_t|$ by the variable:
\begin{eqnarray}
|p_t|=\left[\epsilon+50 \cdot \ln\left(1+\frac{1+t}{1-t}\right)\right]\,\mathrm {MeV},
\end{eqnarray}
where $\epsilon$ is a small parameter and is introduced to avoid divergence in numerical calculations and $t$ varies from -1 to 1. We then discretize Eqs.~(\ref{BSbf11.1}) and (\ref{BSbf21.1}) into $n$ pieces ($n$ is large enough) through the Gauss quadrature rule. The Bethe-Salpeter amplitude can be written as $n$-dimension vectors, $f_{1(2)}^{(n)}$. The coupled integral equations become two matrix equations $f_{1(2)}^{n}=A_{1(2)1}^{n\times n}\cdot f_1^{n}+A_{1(2)2}^{n\times n}\cdot f_2^{n}$ [$A$ corresponds to the coefficients in Eqs.~(\ref{BSbf11.1}) and (\ref{BSbf21.1})]. One can obtain the numerical results of the Bethe-Salpeter amplitude by solving the eigenvalue equation obtained from the above two matrix equations.

In our calculation, we take the values of the parameters as $m_1=938$\,MeV, $m_2=493$\,MeV \cite{K.A.}. According to ChPT, the isospin coefficient $C_{\mathrm {if}}=2$ in the $K^-p\to K^-p$ coupling process \cite{Oset:1997it,Hyodo:2007jq}. For kaons in the meson-meson interaction, $f_K=1.19f_\pi$ ($f_\pi=93$\,MeV) and we should expect similar value here \cite{Ikeda:2012au,Oset:1997it,Hyodo:2007jq}. It can be seen from Eqs.~(\ref{BSbf11.1}) and (\ref{BSbf21.1}) that there is only one free parameter in our model, the mass $M$ of the possible bound state. We vary $M$ from 1300\,MeV to 1450\,MeV in our calculation and find that the nontrivial solution of the eigenvalue equation exists when $M=1422$\,MeV. In other words, the proton and antikaon could form a bound state in this region and its mass is 1422\,MeV. The corresponding numerical results of the Lorentz-scalar functions in the normalized Bethe-Salpeter amplitude, $\tilde{f}_1(|p_t|)$ and $\tilde{f}_2(| p_t|)$, are given in Fig.~\ref{figBSWave}. One should note that the units of $\tilde{f}_1(| p_t|)$ and $\tilde{f}_2(| p_t|)$ are 1 and $\mathrm{MeV}^{-1}$, respectively. In the following, we will take $M$, $\tilde{f}_1(| p_t|)$ and $\tilde{f}_2(| p_t|)$ as input when calculate the decay width of the bound state.
\begin{table*}[tb]
\renewcommand\arraystretch{1.5}
\centering
\caption{The pole position (width) of the $\Lambda(1405)$ in the chiral unitary approach including the Weinberg-Tomozawa term (WT), Weinberg-Tomozawa and Born terms (WTB), and next-to-leading-order (NLO) interaction of ChPT, respectively, within the SIDDHARTA experiment constraints. \cite{K.A.}\label{Table1}}
\begin{tabular*}{\textwidth}{@{\extracolsep{\fill}}lcccccc}
\hline
\multirow{2}*{Approach}         & \multicolumn{3}{c} {Pole 1~(MeV)}& \multicolumn{3}{c}{ pole 2~[MeV]} \\
 &  WT      &WTB & NLO  &  WT      &WTB & NLO \\
\hline
Refs.~\cite{Ikeda:2011pi,Ikeda:2012au}&  1422[16] & 1421[17]&1424[26] &1384[90] & 1385[105]&1381[81]\\
Ref.~\cite{Guo:2012vv}, Fit II  &  -&- & 1421[19] &-  & - &\, 1388[114]\\
Ref.~\cite{Mai:2014xna}, solution\#2& - & - &1434[10]&- & - &1330[56]\\
Ref.~\cite{Mai:2014xna}, solution\#4&  -& - &1429[12]&- & - &1325[90]\\
\hline
\hline
\end{tabular*}
\end{table*}

Then, we apply the numerical solution of the Bethe-Salpeter amplitude to calculate the decay width of $\Lambda^*\to \Sigma^+\pi^-$. We use the following input parameters \cite{K.A.,Oset:1997it,Hyodo:2007jq}: $m_a$=1189\,MeV, $m_b=139$\,MeV, and $C_{\mathrm {if}}=1$. With the parameters determined above, the decay width of the process $\Lambda^*\to\Sigma^+\pi^-$ in our calculation is 15\,MeV.

Several studies point out the existence of the two-pole structure in the region of the $\Lambda(1405)$. The main component may be the $K^-p$ bound state which is narrow and stable and the other is the $\Sigma \pi$ continuum (or scattering) state \cite{K.A.,Ikeda:2011pi,Ikeda:2012au,Guo:2012vv,Mai:2014xna}. The results of the pole structure in the unitary chiral approach within the SIDDHARTA experiment constraints are displayed in Table \ref{Table1}. One can see that our result is in agreement with the pole 1 which is just below the $K^-p$ threshold. This situation supports the existence of the $K^-p$ bound state. Furthermore, according to the PDG, the peak and width of the $\Lambda(1405)$ resonance is $1405.1^{+1.3}_{-1.0}\,$ and $50.5\pm2.0$\,MeV, respectively \cite{K.A.}. We can see the bound state in our calculations is located in the range of the $\Lambda(1405)$ and its decay width is quite smaller than that of $\Lambda(1405)$. That is to say, the $K^-p$ bound state does exist and could contribute to the observed $\Lambda(1405)$, but there may be some other structures in the observed resonance region.

\section{Summary and discussion}
The Bethe-Salpeter formalism has been successfully applied in many theoretical studies concerning heavy mesons, heavy baryons, and molecular bound states automatically including relativistic corrections. In this paper, we studied the possible $s$-wave molecular bound state of the $K^-p$ system in this formalism. Considering the interaction kernel based on the ChPT at the leading order, we established the Bethe-Salpeter equation for the $K^-p$ system in the ladder and instantaneous approximations. Then, we discretized the integral equations and solved the eigenvalue equation numerically. We confirmed the existence of the $s$-wave $K^-p$ bound state in this formalism and obtained its Bethe-Salpeter amplitude. We also calculated the decay width of the $\Lambda^*\to \Sigma^+\pi^-$ process by using the Bethe-Salpeter amplitude. According to our calculation, the mass of the $K^-p$ bound state is compatible to that of the $\Lambda(1405)$ resonance.

In this work, we used the so-called ladder approximation. One may wonder if this approximation is a good one since higher-order graphs could give more important contributions than the ladder graphs. In fact, the legitimacy of the application of the ladder approximation in the Bethe-Salpeter formalism has been studied \cite{Guo:2007mm,Gross:1982nz,CI}. It was shown that including only the ladder graphs in the scalar-scalar system cannot lead to the correct one-body limit \cite{Gross:1982nz} and gauge invariance cannot be maintained within the ladder approximation. To solve these problems, the crossed-ladder graphs should be included at least \cite{Gross:1982nz,CI}. However, in our case, the interaction terms at lowest order $\mathcal O(p)$ of ChPT, which can leads to crossed-ladder graphs, the Born terms, mainly contribute to the $p$-wave interaction \cite{Weingerg:1996}. So, we can adopt the ladder approximation legitimately in our model. Another approximation we took is the instantaneous approximation. In this approximation, the energy exchange between the constituents is neglected. The binding energy of the bound state can be defined as $E_b=M-(m_1+m_2)$. In our calculation, the binding energy is --9\,MeV. This shows that the binding of the constituent particles is weak; hence, the exchange of energy between them can be neglected.

\section{ACKNOWLEDGEMENTs}

This work was supported by the Fundamental Research Funds for the Central Universities (Project No.~3102017OQD052), NSFC-Yunnan United Fund (Project No.~U1302267), National Natural Science Foundation of China (Projects No.~11275025, 11575023, and 11775024), and the National Science Fund for Distinguished Young Scholars (Project No.~31325005).


\begin{thebibliography}{}


\bibitem{K.A.}K.A.~Olive et al.~[PDG Collaboration],
Chin.\ Phys.\ C {\bf 40}, 100001 (2016).


\bibitem{Guo:2017jvc}
  F.K.~Guo, C.~Hanhart, U.G.~Mei\ss ner, Q.~Wang, Q.~Zhao and B.S.~Zou,
  arXiv:1705.00141 [hep-ph].





\bibitem{Hyodo:2011ur}
  T.~Hyodo and D.~Jido,
  Prog.\ Part.\ Nucl.\ Phys.\  {\bf 67}, 55 (2012).



\bibitem{Dalitz:1959dn}
  R.H.~Dalitz and S.F.~Tuan,
  Phys.\ Rev.\ Lett.\  {\bf 2}, 425 (1959).


\bibitem{Dalitz:1960du}
  R.H.~Dalitz and S.F.~Tuan,
  Annals Phys.\  {\bf 10}, 307 (1960).

\bibitem{Alston:1961zzd}
  M.H.~Alston et al.~,
  Phys.\ Rev.\ Lett.\  {\bf 6}, 698 (1961).

\bibitem{Dalitz:1967fp}
  R.H.~Dalitz, T.C.~Wong and G.~Rajasekaran,
  Phys.\ Rev.\  {\bf 153}, 1617 (1967). 


\bibitem{Bohm:1977mz}
  A.~Bohm and R.B.~Teese,
   Phys.\ Rev.\ D {\bf 18}, 4178 (1978).




\bibitem{Hall:2014uca}
  J.M.M.~Hall, W.~Kamleh, D.B.~Leinweber, B.J.~Menadue, B.J.~Owen, A.W.~Thomas and R.D.~Young,
  Phys.\ Rev.\ Lett.\  {\bf 114}, 132002 (2015).

\bibitem{Jido:2003cb}
  D.~Jido, J.A.~Oller, E.~Oset, A.~Ramos and U.G.~Mei\ss ner,
  Nucl.\ Phys.\ A {\bf 725}, 181 (2003).

\bibitem{Bazzi:2011zj}
  M.~Bazzi et al. [SIDDHARTA Collaboration],
  Phys.\ Lett.\ B {\bf 704}, 113 (2011).
\bibitem{Bazzi:2012eq}
  M.~Bazzi et al. [SIDDHARTA Collaboration],
  Nucl.\ Phys.\ A {\bf 881}, 88 (2012).




\bibitem{Ikeda:2011pi}
  Y.~Ikeda, T.~Hyodo and W.~Weise,
  Phys.\ Lett.\ B {\bf 706}, 63 (2011).

\bibitem{Ikeda:2012au}
  Y.~Ikeda, T.~Hyodo and W.~Weise,
  Nucl.\ Phys.\ A {\bf 881}, 98 (2012).

\bibitem{Mai:2012dt}
  M.~Mai and U.G.~Mei\ss ner,
  Nucl.\ Phys.\ A {\bf 900}, 51  (2013).

\bibitem{Guo:2012vv}
  Z.-H.~Guo and J.A.~Oller,
  Phys.\ Rev.\ C {\bf 87}, 035202 (2013).

\bibitem{Mai:2014xna}
  M.~Mai and U.G.~Mei\ss ner,
  Eur.\ Phys.\ J.\ A {\bf 51}, 30 (2015).



\bibitem{Gasser:1984gg}
  J.~Gasser and H.~Leutwyler,
  Nucl.\ Phys.\ B {\bf 250}, 465 (1985).


\bibitem{Ecker:1994gg}
  G.~Ecker,
  Prog.\ Part.\ Nucl.\ Phys.\  {\bf 35}, 1 (1995).

\bibitem{Oset:1997it}
  E.~Oset and A.~Ramos,
  Nucl.\ Phys.\ A {\bf 635}, 99 (1998).






\bibitem{Feynman:1949hz}
  R.P.~Feynman,
  Phys.\ Rev.\  {\bf 76}, 749 (1949).

\bibitem{Salpeter:1951sz}
  E.E.~Salpeter and H.~A.~Bethe,
  Phys.\ Rev.\  {\bf 84}, 1232 (1951).

\bibitem{DL} D. Lurie, Particles and Fields, Interscience Publishers,
New York, 1968, Chap. 9.


\bibitem{Oller:1997ti}
  J.~A.~Oller and E.~Oset,
  Nucl.\ Phys.\ A {\bf 620}, 438 (1997);
  \emph{Erratum:} [Nucl.\ Phys.\ A {\bf 652}, 407 (1999)].



\bibitem{Guo:1996jj}
  X.-H.~Guo and T.~Muta,
  Phys.\ Rev.\ D {\bf 54}, 4629 (1996).  

\bibitem{Guo:2007mm}
  X.-H.~Guo and X.-H.~Wu,
  Phys.\ Rev.\ D {\bf 76}, 056004 (2007).

\bibitem{Xie:2010zza}
  Z.-X.~Xie, G.-Q.~Feng and X.-H.~Guo,
  Phys.\ Rev.\ D {\bf 81}, 036014 (2010).


\bibitem{Feng:2011zzb}
  G.-Q.~Feng, Z.-X.~Xie and X.-H.~Guo,
  Phys.\ Rev.\ D {\bf 83}, 016003 (2011).


\bibitem{Zhang:2013gqa}
  L.~Zhang and X.-H.~Guo,
  Phys.\ Rev.\ D {\bf 87}, 076013 (2013). 


\bibitem{Liu:2015qfa}
  Y.~Liu, X.-H.~Guo and C.~Wang,
  Phys.\ Rev.\ D {\bf 91}, 016006 (2015).  

\bibitem{Liu:2016wzh}
  L.-L.~Liu, C.~Wang, Y.~Liu and X.-H.~Guo,
  Phys.\ Rev.\ D {\bf 95}, 054001 (2017). 






\bibitem{Weingerg:1996}
  S.~Weingerg,
  The quantum theory of the field, in: modern applications, Vol.~2, Cambridge University Press, London, 1996, Chap.~19.

\bibitem{Dalitz:1991sq}
  R.H.~Dalitz and A.~Deloff,
  J.\ Phys.\ G {\bf 17}, 289 (1991).

\bibitem{Hyodo:2007jq}
  T.~Hyodo and W.~Weise,
  Phys.\ Rev.\ C {\bf 77}, 035204 (2008).


\bibitem{Gross:1982nz}
  F.~Gross,
  Phys.\ Rev.\ C {\bf 26}, 2203 (1982).

\bibitem{CI} C.C.~Itzykson and J.-B.~Zuber, Quantum field theory, Mcgraw-Hill,
New York, 1985, Vol.~II, Chap.~10.




\end{thebibliography}

\end{document}